\documentclass[%
 aip,
 apl,
 amsmath,amssymb,
 reprint,%
]{revtex4-1}

\usepackage{CJK}
\usepackage{graphicx}
\usepackage{dcolumn}
\usepackage{bm}
\usepackage{siunitx}
\usepackage{multirow}
\usepackage{subcaption}
\DeclareSIUnit\sq{\ensuremath{\Box}}
\usepackage[colorlinks]{hyperref}

\begin{document}
\begin{CJK}{UTF8}{}
\preprint{AIP/123-QED}

\title{Understanding and minimizing resonance frequency deviations on a 4-inch kilo-pixel kinetic inductance detector array}

\author{S. Shu}
\email{shiboshu@caltech.edu}
\thanks{Current address: California Institute of Technology, Pasadena, California 91125, USA}
\affiliation{ 
Institut de RadioAstronomie Millim\'{e}trique, 38406 Saint Martin d'H\`{e}res, France
}

\author{M. Calvo}
\affiliation{
Institut N\'{e}el, CNRS and Universit\'{e} Grenoble Alpes, 38042 Grenoble, France
}
\author{J. Goupy}
 \affiliation{ 
Institut de RadioAstronomie Millim\'{e}trique, 38406 Saint Martin d'H\`{e}res, France
}
\affiliation{
Institut N\'{e}el, CNRS and Universit\'{e} Grenoble Alpes, 38042 Grenoble, France
}

\author{S. Leclercq}
\affiliation{ 
Institut de RadioAstronomie Millim\'{e}trique, 38406 Saint Martin d'H\`{e}res, France
}

\author{A. Catalano}
\affiliation{
Institut N\'{e}el, CNRS and Universit\'{e} Grenoble Alpes, 38042 Grenoble, France
}
\affiliation{
LPSC, CNRS and Universit\'{e} Grenoble Alpes, 38026 Grenoble, France
}

\author{A. Bideaud}
\author{\\A. Monfardini}
\affiliation{
Institut N\'{e}el, CNRS and Universit\'{e} Grenoble Alpes, 38042 Grenoble, France
}

\author{E.F.C. Driessen}
\affiliation{ 
Institut de RadioAstronomie Millim\'{e}trique, 38406 Saint Martin d'H\`{e}res, France
}

\date{\today}

\begin{abstract}

One of the advantages of kinetic inductance detectors is their intrinsic frequency domain multiplexing capability. However, fabrication imperfections usually give rise to resonance frequency deviations, which create frequency collision and limit the array yield. Here we study the resonance frequency deviation of a 4-inch kilo-pixel lumped-element kinetic inductance detector (LEKID) array using optical mapping. Using the measured resonator dimensions and film thickness, the fractional deviation can be explained within $\pm 25\times 10^{-3}$, whereas the residual deviation is due to variation of electric film properties. Using the capacitor trimming technique, the fractional deviation is decreased by a factor of 14. The yield of the trimming process is found to be 97\%. The mapping yield, measured under a 110~K background, is improved from 69\% to 76\%, which can be further improved to 81\% after updating our readout system. With the improvement in yield, the capacitor trimming technique may benefit future large-format LEKID arrays. 

\end{abstract}


\maketitle
\end{CJK}

Lumped element kinetic inductance detectors (LEKIDs) have been widely developed for astronomical observations~\cite{Day:2003a,Doyle:2008a,Adam:2018a,Wandui:2020aTKID,Austermann:2018a,Wheeler:2018aSuperspec,Walter:2020aSCExAO} in the last decade. Their intrinsic frequency multiplexing property makes LEKIDs suitable for large detector arrays. In practice, the multiplexing factor per readout line depends on the readout bandwidth, resonance frequency spacing, and resonance width, which is usually limited by background radiation for ground-based observations. As the number of pixels per feedline increases, resonance frequency collision between adjacent resonators becomes problematic. When a collision happens, the readout tone may pick up signal from a nearby resonance. This affects the number of pixels that are useful for astronomical observations. Frequency collision is usually caused by the variation of material parameters, such as the film thickness, the superconducting transition temperature, and the resonator dimensions across the wafer. Aluminium is commonly used in LEKIDs design for its easy fabrication, low gap frequency and long quasiparticle lifetime~\cite{Mazin:2020material}. Its low kinetic inductance and resistivity, however, impose thin films ($<80$~nm) and narrow linewidths ($<4$~$\mathrm{\mu m}$), in order to optimize sensitivity and optical coupling. For example, the NIKA2 260~GHz array uses a 18~nm thick Al film with a 4~$\mathrm{\mu m}$ inductor width\cite{Adam:2018a}. Intrinsic variation in these parameters gives rise to a large, uncontrolled deviation of the resonance frequency from the design values, leading to many frequency collisions. This causes the number of functional pixels of the NIKA2 260~GHz array to decrease from 84\% to 70\% under sky background illumination~\cite{Adam:2018a}.

Recently, a corrective technique has been developed, in which scatter in the resonance frequencies is mitigated by a second lithography step adapting the resonator capacitor~\cite{Liu:2017trimming,Shu:2018apl}. This technique has been deployed on LEKID arrays consisting of $\sim 100$ resonators. It has been shown that the deviation from design resonance frequency can be improved by a factor 10, going down from the few percent level to $<1\%$. Array yield has been shown to go up to 97\% on a 1-inch 112-pixel array~\cite{Shu:2018apl}.

In this letter, we take this technique one step further, to a telescope-class, 4-inch kilo-pixel LEKID array. This array is designed for the 260~GHz band of the NIKA2 instrument with 2392 pixels, designed to have their resonance frequencies spaced by only 1.6~MHz. We have analysed the causes for resonance frequency deviation for 920 pixels from this array, and have applied our corrective technique to these pixels. We show that $\sim 90\%$ of the initial frequency deviation can be explained by geometric variations of the individual pixels, and that the technique is ready to be deployed on actual telescope-class arrays.

\begin{figure*}
    \includegraphics[width=0.95\textwidth]{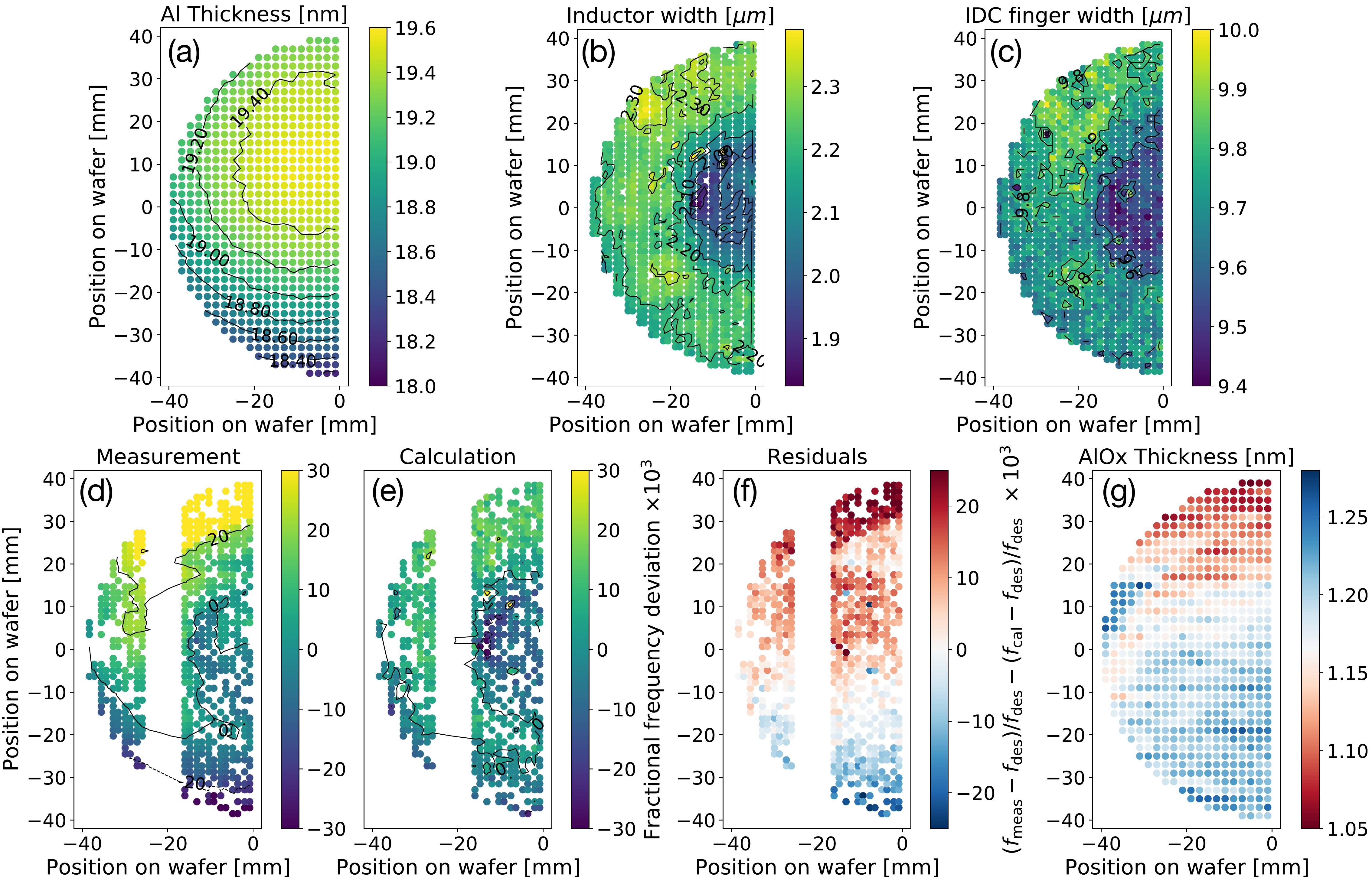}
    \caption{(a) Al thickness map measured using ellipsometer. (b) Inductor width map measured using SEM. (c) Capacitor width map measured using SEM. (d) The measured fractional frequency deviation before trimming. (e) The calculated fractional frequency deviation. (f)The difference of $\delta f/f$ between (d) and (e). (g) AlOx thickness map measured using ellipsometer.}
\label{fig:dfof}
\end{figure*}

The detector array design in this study is based on the current NIKA2 \SI{260}{\GHz} array~\cite{Calvo:2016a}. To optimally take advantage of the angular resolution possible at the IRAM 30-m telescope, the meandering inductor is reduced to a surface area of $1\times 1~\mathrm{mm}^2$~\cite{Shu:2018jltp}. To fully cover the 80~mm diameter focal plane, the total number of pixels is increased to 2392, distributed over 8 feedlines. The resonators are made of $t=18$~nm thick Al on a \SI{250}{\um}-thick high-resistivity silicon substrate. In our design and calculations, we assumed a surface resistance $R_s=\SI{1.6}{\ohm\per\sq}$ and a surface inductance $L_s=\SI{2}{\pico\henry\per\sq}$. A 200~nm thick Al film is deposited on the back side of the wafer, serving as the ground of the microstrip feedline and the backshort for optical coupling. The inductor is shaped in a 3rd-order Hilbert curve with a width of $2.5~\mathrm{\mu m}$. An 8-finger interdigitated capacitor (IDC) is used to tune the resonance frequencies from \SIrange{1.85}{2.35}{\GHz}, by decreasing the IDC finger lengths pair by pair. The geometry detail is shown in the supplementary material. On different feedlines, the number of resonators varies from 276 to 332 and the designed frequency spacing ranges from 1.51~MHz to 1.82~MHz. The resonance frequency shift due to coupling between adjacent resonators is optimized to be $<$55~kHz. The maximum crosstalk between feedlines is smaller than \SI{-21}{\dB}. The coupling quality factor $Q_c$ is optimized to \num{1e4} to match the expected internal quality factor under a \SI{50}{\K} sky radiation. The array design detail is shown in the supplementary material.

After fabrication, dicing and wire bonding, the array is cooled down to 100~mK in a dilution cryostat. Due to a limited number of connections in our test cryostat, only 4 out of 8 feedlines were connected. One feedline was found to be broken, the resulting measurements were performed on 3 feedlines with 920 designed resonators. Both S21 and mapping measurement were performed with the cryostat window open. To locate the position of the resonances on wafer, we performed an optical mapping on this array using a beam mapper~\cite{Shu:2018apl, Shu:2018jltp}. This mapper consists of a blackbody cooled down to 50~K, in front of which a moving 300~K metal ball acts as a source. The Teflon window of this mapper has a 20\% emissivity at 300~K, which increases the total background radiation to the equivalent of a 110~K blackbody. The mapping was performed by scanning y-axis at fixed $x$ position. The $x$ step is 1~mm, corresponding to 0.4~mm on focal plane. We scan the metal ball at \SI{4}{\mm\per\second}, corresponding to \SI{1.6}{\mm\per\second} on focal plane. With a readout frequency of 23~Hz, this gives a position resolution of \SI{0.17}{mm}. Compared with the designed 1.4~mm pitch size between resonators, this configuration has enough accuracy to locate them individually. On each feedline, the resonators were measured simultaneously using a NIKEL readout board~\cite{Bourrion:2016a}.  Fig.~\ref{fig:dfof}d shows the measured fractional resonance frequency deviation, defined as $\delta f/f=(f_{\textrm{meas}}-f_{\textrm{des}})/f_{\textrm{des}}$, where $f_{\textrm{des}}$ is the initially designed resonance frequency and $f_{\textrm{meas}}$ is the measured resonance frequency. 

To understand the cause of this frequency deviation, we calculated the frequency deviation through equation $f_{\textrm{cal}}= 1/(2\pi\sqrt{(L_m+L_k)C})$, with $L_m$ the magnetic inductance, $L_k$ the kinetic inductance, and $C$ the capacitance. From simulation using a commercial software package (Sonnet), we have $L_m = 10.56  (w/w_0)^{-0.16}$ nH with the designed inductor width $w_0=2.5 \mathrm{\mu m}$, and $C = C_0(w_c/w_{c0})^{0.765}$ with $C_0 = 1/(2\pi f_0)^2L_m$, where $C_0$ is the capacitance for each resonator with the designed IDC finger width $w_{c0}=10\mathrm{\mu m}$ and $f_0$ is the resonance frequency with $L_k=0$. The kinetic inductance is given by $L_k=L_s (l/w)$, where $l=9$ mm is the inductor length. The thickness dependence of $L_s$ is non-trivial since both critical temperature and resistivity vary significantly with film thickness in our thickness range. We use an empirical relation $L_s = 306t^{-1.67}$ pH/$\sq$, with $t$ in nanometers, which is obtained from a series of independent measurements on sputter deposited films, while noting that this relation might not be perfectly adapted for our current e-beam evaporated film.

The geometry dimensions used in calculation are shown in Figure~\ref{fig:dfof}a-c. The aluminium film thickness was measured just after deposition, using ellipsometry. The thickness varies from 18.0~nm to 19.6~nm over the wafer, in a concentric fashion that is expected from thin film deposition using e-beam evaporation. The inductor width $w$ and IDC finger width $w_c$ are measured for each individual pixel using scanning electron microscopy (SEM). Both dimensions show a large variation, with notably a large size reduction in the center of the array. These variations are possibly due to variations in the photolithography process, like the variation in resist thickness, or variations in local wet etching speed, and are currently under investigation.

Using the calculation described above, we calculate the explained fractional frequency deviation, as shown in Fig.~\ref{fig:dfof}e. The residual, unexplained, frequency deviation is shown in Fig.~\ref{fig:dfof}f. The average residual frequency deviation is $5\times 10^{-3}$, with a standard deviation of $10\times 10^{-3}$ ($\sim$20~MHz). This is not sufficiently precise to predict resonance frequencies without an optical measurement. We note that the residuals are most significant in a small area in the upper left region of the wafer, which region is also observed in the results after trimming. Moreover, we find a strong correlation ($r=0.8$) between residual frequency deviation and aluminum oxide (AlOx) thickness (measured with ellipsometry before film processing, and shown in Fig.~\ref{fig:dfof}g) for the region where the residuals are $<15\times 10^{-3}$, suggesting that the residuals are mainly related with $L_s$. We have no means of measuring $L_s$ or other electrical film parameters locally, due to the dense packing of this telescope-class array.

In the optical mapping, the positions of 635 (69\%) resonators were identified. For the unidentified resonators, we used radial basis functions to interpolate or extrapolate their resonance frequencies. We re-designed the resonance frequencies after trimming as $f_{\textrm{redes}}$, ranging from \SIrange{1.96}{2.46}{\GHz} for each feedline. As both $f_{\textrm{des}}$ and $f_{\textrm{redes}}$ use the same relation between IDC finger lengths and resonance frequencies, the corresponding trimming frequencies of the patterns on the trimming mask are given by $f_{\textrm{trim}} = f_{\textrm{des}}f_{\textrm{redes}}/f_{\textrm{meas}}$, assuming the same frequency deviation
\begin{equation}
    \frac{\delta f}{f} =  \frac{f_{\textrm{redes}}-f_{\textrm{trim}}}{f_{\textrm{trim}}}.
\end{equation}
The final trimming mask is defined by the difference of the IDC patterns between $f_{\textrm{des}}$ and $f_{\textrm{trim}}$ designs. 

As described before~\cite{Shu:2018apl}, the trimming process consists of a contact lithography and Al wet-etching step. As the photoresist residue in small patterns can block wet-etching, an oxygen plasma was applied to remove the residue after development. During etching process, a large part of the backside 200~nm Al was etched away, so a new layer of 200~nm Al was sputtered on top of the Al residue.

\begin{figure}
\begin{center}
\includegraphics[width=3.2in]{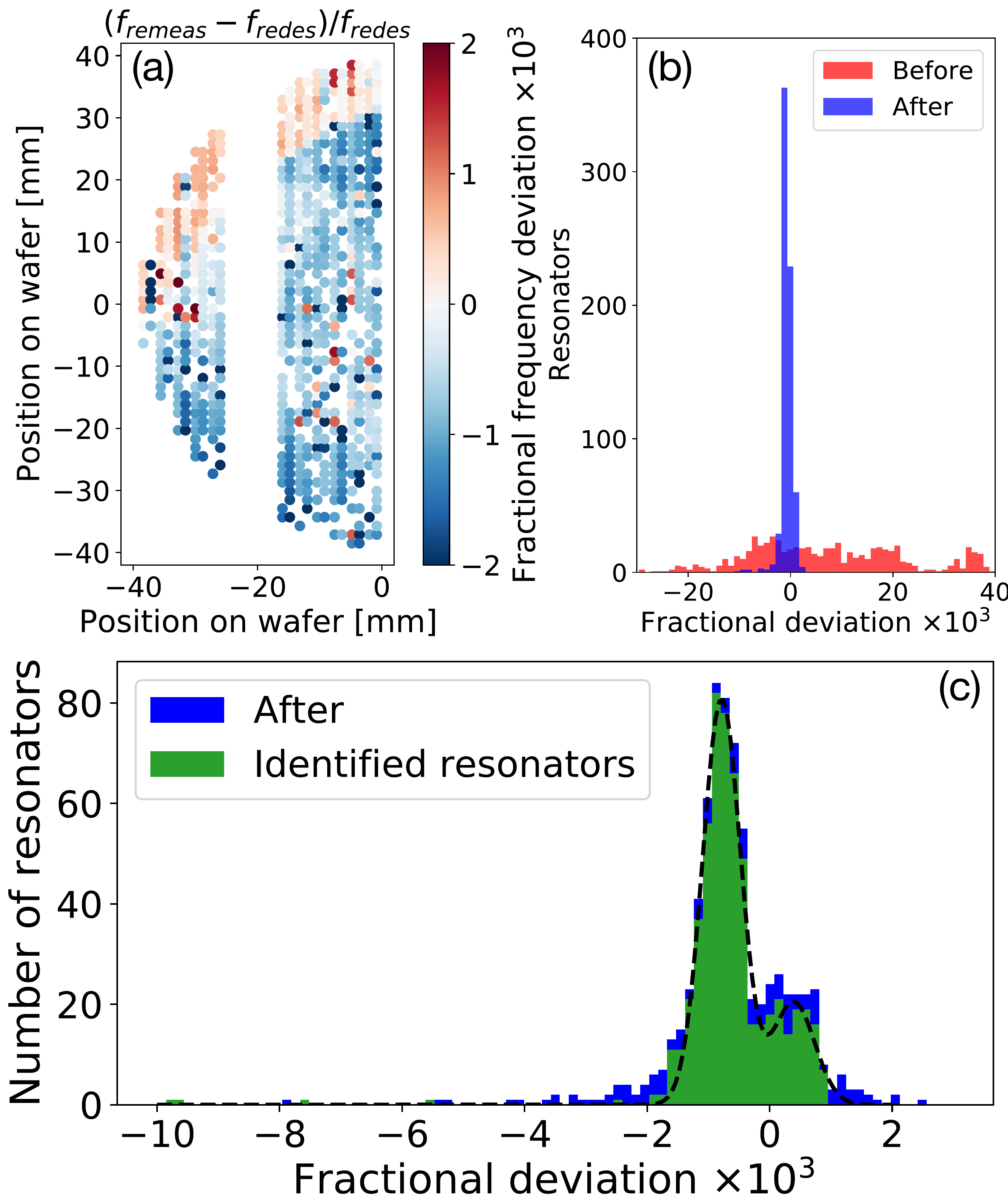}
\caption{(a) $\delta f/f$ map after trimming. (b) Histograms of $\delta f/f$ before and after trimming. (c) Comparison of the trimming results on all resonators and the resonance frequencies identified resonators. The difference suggests the inter/extrapolation gives a similar result. A two Gaussian fit of $\delta f/f$ of the identified resonators is plotted (dashed line), which gives $\mu_1 = \num{-7.8e-4}$ and $\sigma_1=\num{3.1e-4}$ and $\mu_2 = \num{3.9e-4}$ and $\sigma_2 = \num{3.4e-4}$.}
\label{fig:dfof_af}
\end{center}
\end{figure}

After trimming, the array was re-bonded and cooled down to 100~mK using the same setup as in the first characterization. S21 and mapping measurements are applied on the same three feedlines. The resonance frequencies after trimming $f_{\textrm{remeas}}$ range as the redesigned frequency $f_{\textrm{redes}}$. After trimming, the overall standard deviation of $\delta f/f$ is $1.13\times 10^{-3}$, improved by a factor of 14. For the resonators identified beforehand, the standard deviation is $\delta f/f$ is $0.96\times 10^{-3}$, indicating that the radial basis function used to inter/extrapolate the resonance frequencies works well on large format arrays. We also notice that trimming one or two pairs of IDC fingers give similar accuracy.

Fig.~\ref{fig:dfof_af}a shows two groups of resonators on the array, with a similar standard deviation of $\sim 3\times 10^{-4}$ but different mean values. One group has $\delta f/f<0$, and its mean value of -1.7~MHz is consistent with previous trimming results~\cite{Liu:2017trimming,Shu:2018apl}, due to Al film aging. The other resonator group with $\delta f/f>0$, is located at the upper region of the array, partly overlapping with where the residuals $>15\times 10^{-3}$. From the residual map (Fig.~\ref{fig:dfof}g) we know that this region has a lower $L_s$ compared with calculation, indicating a lower $R_s$. However, this should not give two groups of $\delta f/f$ if aging happens uniformly. This positive $\delta f/f$ suggests that $L_s$ is either unchanged (no aging) or slightly increased, as $L_s$ should not decrease. By checking previous data, we confirm that this AlOx pattern was reproducible. One possible reason is nonuniformity of the argon plasma cleaning, just before the Al deposition, which creates a varying roughness of the substrate. Larger substrate roughness would facilitate film ageing.

\begin{figure}
\begin{center}
\includegraphics[width=3.2in]{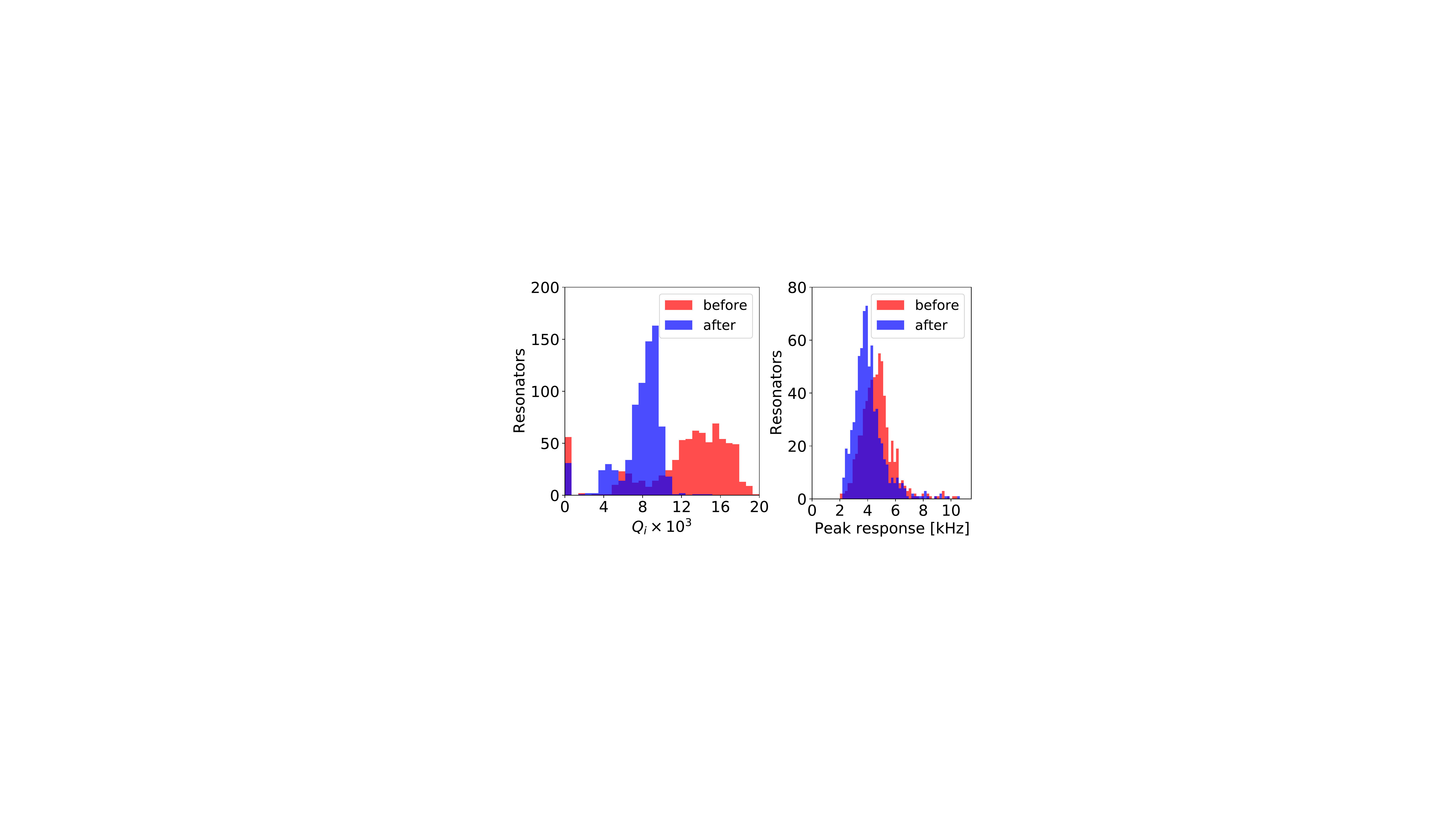}
\caption{(a) Histogram of $Q_i$ under a 110~K blackbody radiation before and after trimming. The resonances shown in the histogram as $Q_i=0$ do not have proper fitting results possibly due to frequency collision. (b) Peak resonator responses of the 300 K source.}
\label{fig:Qi}
\end{center}
\end{figure}

$Q_c$ and $Q_i$ are extracted from S21 measured using VNA under a 110~K background. After trimming, $Q_c\sim 10^4$ remains the same, while $Q_i$ is decreased by \SI{40}{\percent} from $14\times 10^3$ to $8.4\times 10^3$. As using a $40$~K background radiation gives $25\percent$ higher $Q_i$ in both cases, $Q_i$ here are limited by the 110~K background radiation. In a previous experiment, we observed an average dark $Q_i$ of $60\times 10^3$ after trimming, much higher than the optical $Q_i$ here. The film aging and the trimming process may have little effect on our $Q_i$, which is limited by the loss of quasiparticles created by photons. To compare the sensitivity, we directly take the peak response of the mapping source, shown in Fig.~\ref{fig:Qi}b, as the noise is dominated by the $1/f$ noise of the background temperature variation and the readout board. The peak response of the mapping source decreases from \SIrange{4.6}{3.9}{\kHz} after trimming (Fig.~\ref{fig:Qi}), consistent with the 21\% decrease of $Q_r$ from $9.3\times10^3$ to $7.3\times10^3$. 

As the inductor geometry is untouched during trimming and the background temperature of the mapping system is stable, we can remove the influence of dimensions and incident power. The high trimming accuracy of $|\delta f/f|<2\times 10^{-3}$ suggests that the change of sheet resistance $R_s$ and gap energy $\Delta$ of the Al film should also be at the same level, according to $L_s=\hbar R_s/\pi \Delta$, which cannot explain the 40\% degradation of $Q_i$. Using Mattis-Bardeen theory~\cite{Mattis:1958a}, the equivalent temperatures corresponding to the observed $Q_i$ are 334~mK and 361~mK before and after trimming, much larger than the 100~mK bath temperature, suggesting a small change in heat sinking may give this degradation of $Q_i$. The heat sinking of this array is determined by the contact between the backshort Al film and the Al sample holder. In a similar measurement, we observed that using an Al-Au bilayer as a backshort and a copper sample holder increased the $Q_i$ from $3\times10^3$ to $10\times10^3$ under a 300~K optical loading. Because the backside Al is re-deposited, as described above, we suggest that the decrease of $Q_i$ is due to a decrease in heat sinking. This could be remedied by using a Ti-Au bilayer backshort, which can trap phonons~\cite{Karatsu:2019mitigation}, improve heat-sinking, and is resistant to Al etching process.

The yield of this array was measured under the 110~K blackbody radiation, a typical on-telescope condition. For reference, the on-telescope sky background radiation varies from 30~K to 180~K depending on the atmospheric opacity~\cite{Perotto:2020calibration}. After trimming, the yield is increased from 69\% to 76\% counted from optical mapping results. For all counted pixels, no crosstalk is observed from the 2-dimensional mapping results, with signal to noise ratio of 10. 45 resonators are missed by the readout system due to the limited number of readout tones. Including these missed resonators, the final yield is 81\%, 10\% higher than the on-telescope yield of the current NIKA2 260~GHz array~\cite{Adam:2018a}, while the initial fabrication yield is 84\% for both arrays. The yield could be further increased by improving our fabrication process. $23$ resonances were broken during trimming process, giving the trimming fabrication yield of \SI{97}{\percent}. 

In conclusion, we have studied and minimized the resonance frequency deviations in a 4-inch kilo-pixel LEKID array. The calculation agrees with the measurement within an accuracy of $\pm 25\times 10^{-3}$. The decrease of the optical-loaded quality factor, after trimming, could be explained by a degradation of the heat sinking instead of a film property change. After trimming, the mapping yield, measured under a 110~K background, is improved from 69\% to 76\%, which can be further improved to 81\% after updating our readout system. This $7\sim 12\%$ improvement in yield within fixed readout bandwidth suggests that the trimming technique is capable to improve the on-telescope yield, which may benefit future large-format LEKID arrays.

See supplementary material for the design in detail. 

The authors thank D. Billon-Pierron and A. Barbier for experimental help, and K.F. Schuster for useful discussions. This work has been partially funded by LabEx FOCUS ANR-1-LABX-0013.

\section*{Data availability}
The data that support the findings of this study are available from the corresponding author upon reasonable request.

%

\newpage

\section{Supplementary material}
\subsection{Array design}

\begin{figure}
    \centering
    \includegraphics[width=0.8\linewidth]{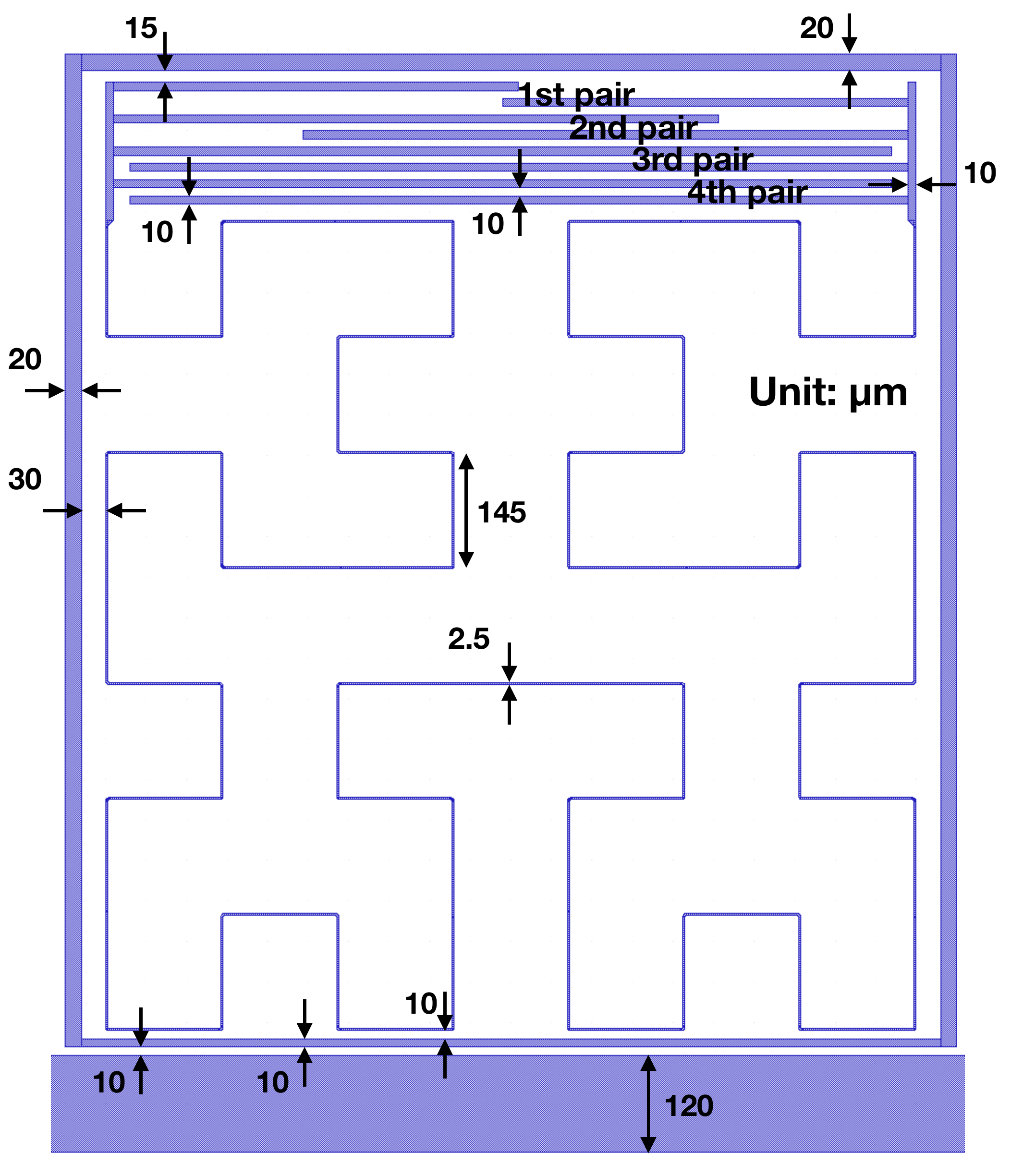}
    \caption{A single LEKID pixel design in detail.}
    \label{fig:single}
\end{figure}

\begin{figure}[h]
    \centering
    \includegraphics[width=0.8\linewidth]{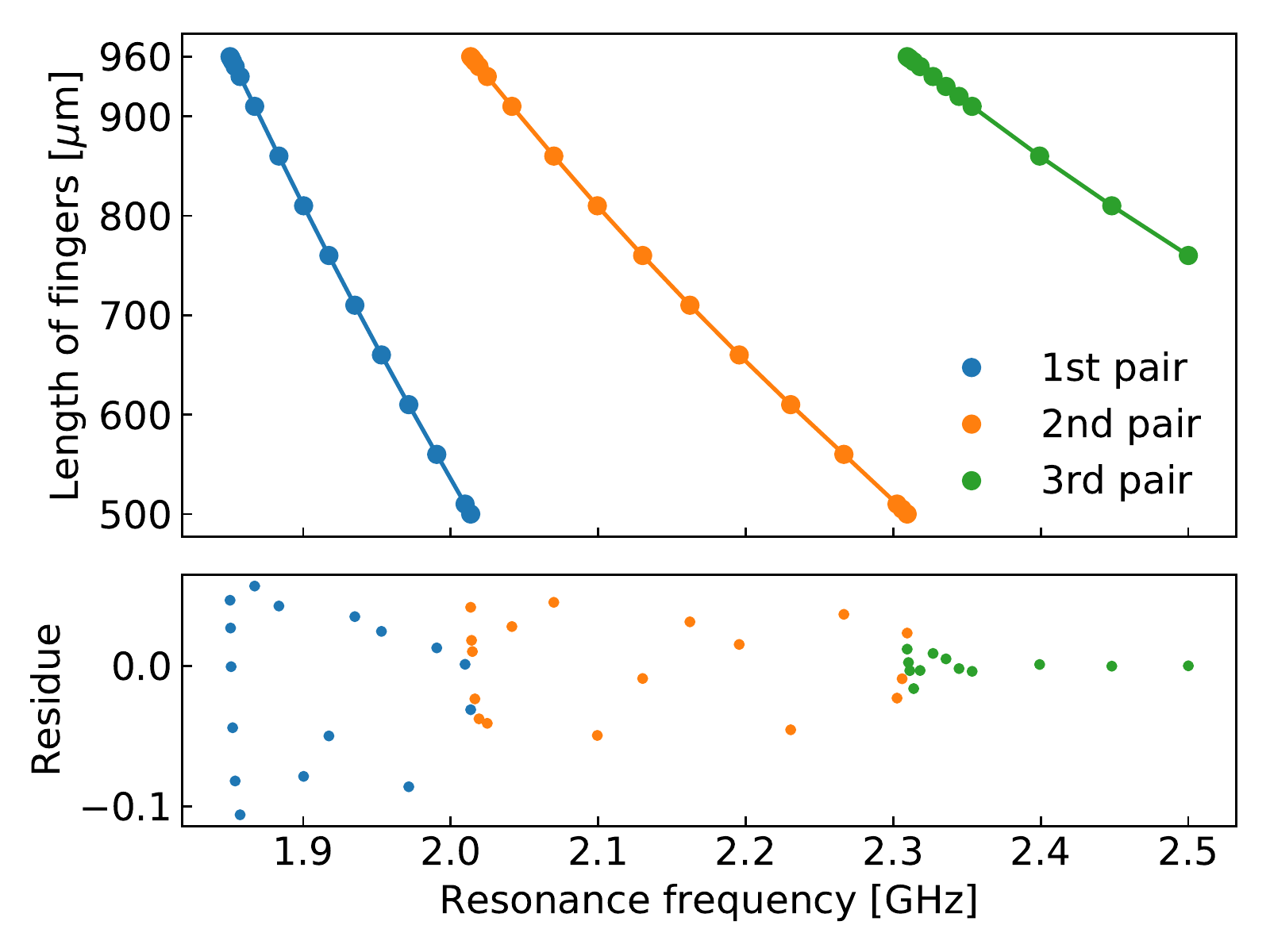}
    \caption{Design of the resonance frequency by changing the length of capacitor fingers. The simulated data is fitted using polynomial functions. The residue is within \SI{100}{\nm}.}
    \label{fig:fres}
\end{figure}

\begin{figure}
    \centering
    \includegraphics[width=0.8\linewidth]{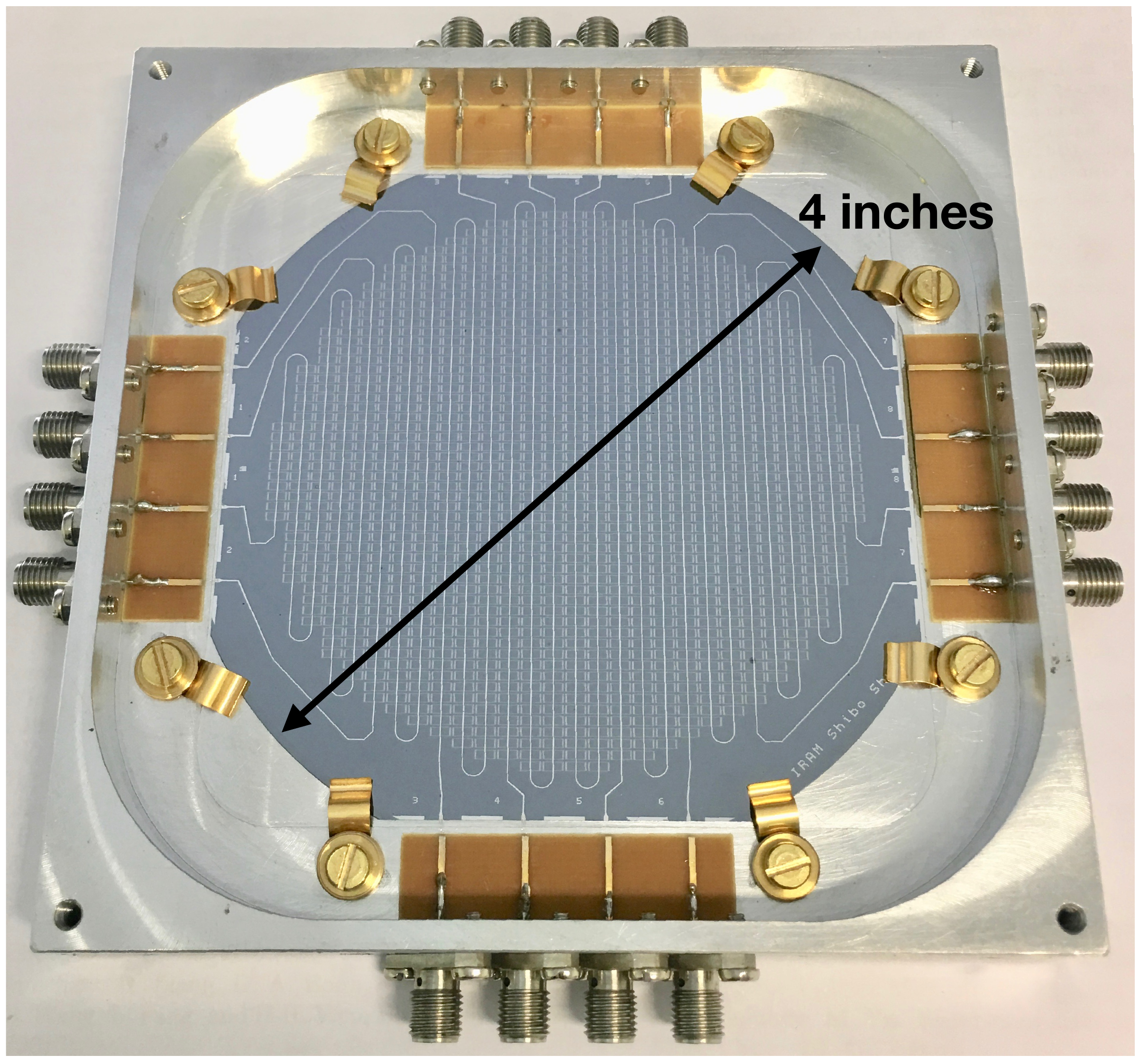}
    \caption{The photo of this 4-inch array in a detector holder.}
    \label{fig:holder}
\end{figure}

The whole focal plane of the NIKA2 instrument is a 80~mm diameter circle. With inductor size of 1~mm, in total there are 2392 resonators on 8 microstrip feedlines. The pitch size is optimized to be 1.4~mm. The single pixel design is shown in Fig.~\ref{fig:single} in detail. Resonance frequencies are tune by shortening the length of IDC fingers pair by pair. The longest finger length is \SI{960}{\um} and the shortest length is \SI{500}{\um}. The resonance frequencies are simulated in Sonnet software with several different IDC finger lengths. A polynomial fit is applied to extracted the relation between finger length and resonance frequencies, shown in Fig.~\ref{fig:fres}. The fitting residue is smaller than $\pm 0.1 \mathrm{\mu m}$. In terms of resonance frequency, this $ 0.1 \mathrm{\mu m}$ corresponds to 5~kHz, much smaller than the simulated $<55$~kHz electrical crosstalk. The resonance frequency after trimming is also defined in this way, so no extra simulations are needed for the re-designed resonance frequencies. The full array after fabrication is shown in Fig.~\ref{fig:holder}.

\end{document}